\documentclass[twocolumn,showpacs,preprintnumbers,amsmath,amssymb]{revtex4}
\usepackage{graphicx}
\usepackage{dcolumn}
\usepackage{bm}

\begin{document}

\preprint{APS}

\title{Majority-vote model on spatially embedded networks:
crossover from mean-field to Ising universality classes}

\author{C. I. N. Sampaio Filho$^1 \footnote{Correspondence to: cesar@fisica.ufc.br}$, T. B. dos Santos$^1$, A. A. Moreira$^1$, F. G. B. Moreira$^2$,  J. S. Andrade Jr.$^{1}$}

\affiliation{$^1$Departamento de F\'{i}sica, Universidade Federal do Cear\'a, 60451-970 Fortaleza, Cear\'a, Brasil\\
$^2$Departamento de F\'{i}sica Te\'{o}rica e  Experimental, Universidade Federal do Rio Grande do Norte, 59072-970, Natal-RN, Brasil}

\date{\today}%

\begin{abstract}
We study through Monte Carlo simulations and finite-size scaling analysis the nonequilibrium phase transitions of the majority-vote model taking place on spatially embedded networks. These structures are built from an underlying regular lattice over which long-range connections are randomly added according to the probability, $P_{ij}\sim{r^{-\alpha}}$ , where $r_{ij}$ is the Manhattan distance between nodes $i$ and $j$, and the exponent  $\alpha$ is a controlling parameter [J. M. Kleinberg, Nature 406, 845 (2000)]. Our results show that the collective behavior of this system exhibits a continuous order-disorder phase transition at a critical parameter, which is a decreasing function of the exponent $\alpha$. Precisely, considering the scaling functions and the critical exponents calculated, we conclude that the system undergoes a crossover among distinct universality classes. For $\alpha\le3$ the critical behavior is described by mean-field exponents, while for $\alpha\ge4$ it belongs to the Ising universality class. Finally, in the region where the crossover occurs, $3<\alpha<4$, the critical exponents are  dependent on $\alpha$.
\end{abstract}

                                                
\maketitle

\section{\label{sec:level1}Introduction} 

It is a remarkable feature of the theory of critical phenomena to condense a large range of systems that undergoes a continuous phase transition in terms of universality classes \cite{lubeckBook2008}. Fundamental properties such as symmetries or dimensionality just can influence the critical exponent, regardless of the microscopic details of interactions. However, under some conditions, it is possible to observe a crossover phenomenon for a given system or model, namely, a change in its universality class. Examples of crossover phenomena are known either in equilibrium, such as in ferromagnetic systems \cite{binderPRE1993,luijtenPRL1996}, as well as in nonequilibrium statistical physics \cite{lubeckPRL2003,lubeckPRE2004,hernanPRL2010,gallosPNAS2012,sampaio2013,sampaioSREP2015}, where the zero law of thermodynamics is not satisfied. In the present work we analyse a crossover from mean-field to Ising universality classes in a nonequilibrium model, namely, the majority-voter model. However, as will be describe, these regimes are separated by a singular region where the critical exponents change continuously. 

The majority-vote model (MVM) with noise \cite{oliveira1991,oliveira1992} is a nonequilibrium model system, which presents up-down symmetry and a continuous order-disorder phase transition. The nature of the transition and the phase diagram for the MVM defined on both regular and complex networks have been extensively investigated, including a significant number of generalizations~\cite{bradyPRE2003,brady2010,sampaio2011,sampaio2013}. Besides their own motivation within the context of nonequilibrium Statistical Mechanics, these studies have focused on the area of phase transitions and critical phenomena to improve our understanding on the robustness and formation of social consensus \cite{galam2002,castellano2009,songScience2010,kitsakNatPhys2010,barthelemyPhysRep2011}.

In the present study, we perform Monte Carlo simulations and employ the finite-size scaling theory to obtain the phase diagram and critical behavior of the the MVM on spatially embedded networks \cite{kleinberg2000,havlinNatPhys2011}, namely, networks constituted of $d$-dimensional lattices as substrates over which long-range connections are randomly added to connect any two sites according to a probability that depends on the distance between these sites. 

The remaining of the paper is organized as follows. In Section II we describe the main features of the social networking model proposed by  Kleinberg and the majority-vote dynamics used here to determine the time evolution of the Ising variables associated to each node of this network. In Section III the results of our simulations are presented and the finite-size scaling analysis is used to investigate the critical properties of the model. We conclude in Section IV.
  
\section{\label{sec:level2}The model definition}

In order to study the effects of nonlocal interactions on the global ordering, we associate one Ising spin variable to each one of the $N$ sites of a regular lattice and perform Monte Carlo simulations considering each spin evolving in time according to the majority-vote dynamics \cite{oliveira1991,oliveira1992}. Here we adopt the Kleinberg's network \cite{kleinberg2000} as a model for introduction long-range couplings between sites in the system. Starting from a regular $d$-dimensional lattice each site is connected with its $2d$ nearest-neighbor sites. Next, every site $i$ can receive a \textit{directed} connection with a site $j$ with probability $P_{ij}{\sim}r_{ij}^{-\alpha}$, where $r_{ij}$ is the Manhattan distance defined by the number of connections separating the nodes $i$ and $j$ in the underlying regular lattice, and $\alpha$ is the parameter that controls the length of these long-range connections (shortcuts). Since the probability of long-range connections depends on the distance between sites, this network are said to be \textit{spatially embedded} \cite{havlinNatPhys2011}. Here, we consider a two-dimensional topology as substrate and the addition of long-range connections \cite{andradePRL2010,andradePRE2013}. 

Previous studies have shown that the MVM model on regular lattices undergoes a typical order-disorder phase transition at the stationary state \cite{oliveira1991,oliveira1992}. Its dynamics is governed by the probabilities $p$ to attribute to a randomly chosen spin the same state of the majority of its neighbors, and $q=1-p$ the opposite state. Therefore, the control parameter $\alpha$ and the probability $q$ define the parameter space for the phase diagram of the MVM on Kleinberg's network. The critical noise approaches a limiting value as $\alpha\to\infty$, but due to the extra connections, remains above the value for a regular square lattice,  $q_{c} = 0.075(1)$~\cite{oliveira1992}.

The MVM is microscopically irreversible, i.e., its stationary state does not satisfy detailed balance and presents up-down symmetry. The former property implies that we are dealing with a nonequilibrium system in which neither energy nor temperature are defined, whereas the existence of up-down symmetry would ensure the Ising universality class for the MVM on regular lattices. Remarkably, we show here that the addition of nonlocal interactions, via the Kleinberg's prescription, modifies this scenario. Indeed, our results show that the universality class of the majority-vote model on these networks do depend on the range of the control parameter $\alpha$. 

In order to study the effect of the noise parameter $q$ and the control parameter $\alpha$ on the phase diagram and critical behavior of the majority-vote model, we consider the magnetization $M_{N}$, the susceptibility $\chi_{N}$, and the Binder's fourth-order cumulant $U_{N}$, which are defined by

\begin{equation}
 M_{N}(q) =  \left< \left<  m  \right>_{time}\right>_{sample}, 
 \label{eq01}
\end{equation}

\begin{equation}
 \chi_{N}(q) = N \left[\left< \left< m^{2} \right>_{time}  - \left< m \right>_{time}^{2} \right>_{sample}\right],
 \label{eq02}
\end{equation}

\begin{equation}
 U_{N}(q) = 1 - \left< \frac{\left< m^{4} \right>_{time}}{3\left< m^2 \right>_{time}^{2}} \right>_{sample},
 \label{eq03}
\end{equation}
where $N$ is the number of spins in the system and $m = |\sum_{i=1}^{N}\sigma_{i}|/N$. The symbols $<\cdots>_{time}$ and $<\cdots>_{sample}$, respectively, denote time averages taken at the stationary state and configurational averages taken over several samples. For a fixed value of $\alpha$, we have performed Monte Carlo simulations on Kleinberg's networks with $N = 2500, 10000, 22500, 40000, 90000, 250000$, and periodic boundary conditions are applied. Time in our simulations is measured in Monte Carlo steps ($MCS$). More precisely, one Monte Carlo step is accomplished when we choose randomly $N$ spins and try to flip each one with the probability rate
\begin{equation}
w(\sigma_{i}) = \frac{1}{2}\left[1 - (1-2q)\sigma_{i}S(\sum_{\delta=1}\sigma_{i+\delta}) \right],
\label{eq04}
\end{equation}
where the summation is over all spins connected with the chosen spin $\sigma_i$, and $S(x) = sgn(x)$ if $x\neq 0$ and $S(0) = 0$ otherwise. We wait $10^{5}$ $MCS$ for the system to reach the steady state and the time averages are calculated based on the next $10^5$ $MCS$. At the critical region, larger runs are performed with $2\times 10^{5}$ $MCS$ to reach the steady state and $10^{6}$ for computing time averages. For all sets of parameters $(q,\alpha)$, at least $100$ independent samples are considered in the calculation of the configurational averages. Moreover, the simulations were performed using different initial spin configurations. 

\section{\label{sec:level3} Results and Discussion}

\begin{figure}[t]
\includegraphics*[width=\columnwidth]{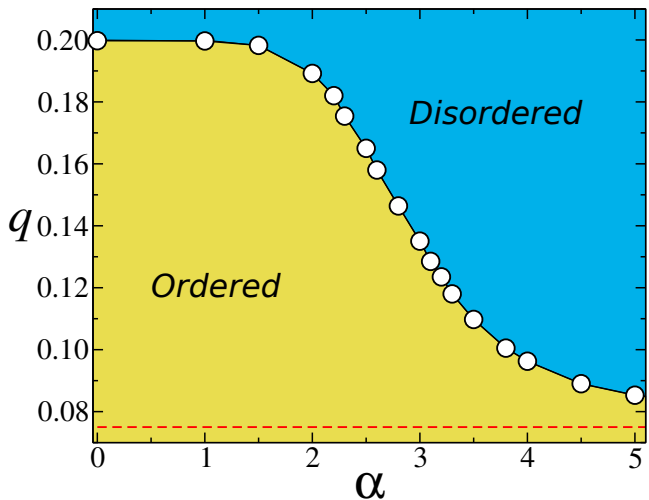}
\caption{(Color online) Phase diagram for the majority-vote model on Kleinberg's networks. The critical parameter $q_{c}$ decreases monotonically with the control parameter $\alpha$. The critical values remain always above the value for the regular square lattice $q_{c}=0.075$ \cite{oliveira1992} (dashed line).}
\label{fig01}
\end{figure}

In the thermodynamic limit $(N\to\infty)$, we expect the system to show nonzero magnetization only below the critical noise $q_{c}(\alpha)$. In Fig.~(\ref{fig01}) we show the phase diagram of the MVM on Kleinberg's networks. For each value of the parameter $\alpha$, the critical value $q_{c}(\alpha)$ is obtained by calculating the Binder's fourth-order cumulant $U_{N}(q)$, Eq.~(\ref{eq03}), as a function of the noise parameter $q$, considering networks with different number of nodes $N$. For sufficiently large system sizes, these curves intercept each other at a single point $U(q_{c})$. Since the Binder's cumulant has zero anomalous dimension \cite{binder1981}, the resulting value of the critical parameter $q_{c}(\alpha)$ is independent of $N$.
As depicted, the phase diagram in the $\alpha \times q$ parameter space of Fig.~\ref{fig01} shows that the critical noise $q_{c}$ decreases monotonically with the control parameter $\alpha$. This reflects the fact that as larger $\alpha$ is shorter is the density of long-range links (shortcuts), and vice-versa. For $\alpha>6$, the curve for $q_{c}(\alpha)$ presents an asymptotic behavior (dashed line) to the value $q_{c}=0.075(1)$ which corresponds to the critical point of the regular square lattice \cite{oliveira1992}. In fact, in the limit $\alpha\to\infty$ we recover the topology of a square lattice and therefore the majority-vote model is described by the $2D$ Ising universality class~\cite{oliveira1992,sampaio2011,sampaio2013}. For $\alpha=0$, however, the system is described by the mean-field theory~\cite{choiPRE2002,lubeckPRL2003,mendes2008}, since for this value of the control parameter the probability of adding a shortcut is the same for all pairs of sites and independent of their distances \cite{newmanPRE1999,wattsNature2000,kleinberg2000}. Our next results characterize this crossover~\cite{lubeckPRL2003,lubeckPRE2004} from the mean-field to the $2D$ Ising universality class.  

\begin{figure}[htb]
\includegraphics*[width=\columnwidth]{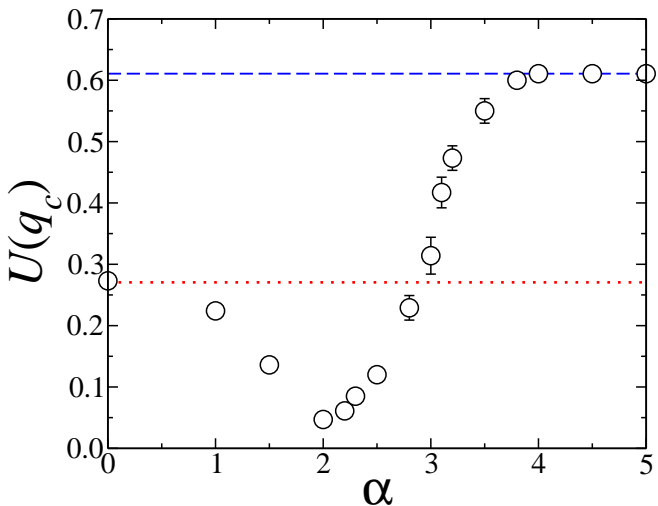}
\caption{(Color online) Binder's cumulant calculated at the critical noise $q_{c}$. The bottom dotted line (red color online) corresponds to the mean-field value ($U^{*} = 0.2705208$) \cite{brezin1985} and the upper dashed line (blue color online) to the limit for the $2D$ Ising universality $U^{*} = 0.61069014(1)$\citep{bloteJPhysA1993,sampaio2011}. For some data points the error bars are smaller than the symbols. A minimum of $U(q_{c})$ at $\alpha=2$ is reminiscent of the optimal navigation condition with local knowledge \cite{kleinberg2000}. The $2D$ Ising regime is recovered for $\alpha \ge 4$.}
\label{fig02}
\end{figure}

In general, some dependency of the critical noise value on the control parameter $\alpha$ should be expected since critical points (for example, critical temperatures \cite{sanchesPRL2002} and critical rates of surviving \cite{dickmanPRE2012})) are not universal properties. Nevertheless, it is remarkable that the value of the Binder's cumulant at the critical noise also depends on $\alpha$, as shown in Fig.~\ref{fig02}. For each value of the parameter $\alpha$ considered, we determine $U^{*}$ (open circles) as the intersection point of the set of curves of the Binder's cumulant $[U_{N}(q)]$. Indeed, for systems with the same symmetry of the Ising model in the regime of short-range interactions, the Binder's cumulant takes the value $U^{*} = 0.61069014(1)$, considering  square lattices with periodic boundary condition \cite{bloteJPhysA1993,salasJSTAT2000,blote2002,selkeJAMath2005,malakisPRE2014}. For the case of the majority-vote model on the same topology, it was found the same value for the critical Binder's cumulant \cite{sampaio2011}. In the mean-field regime, one has the reference value $U^{*} = 0.2705208$ \cite{brezin1985,luijtenPRL1996,luijtenPRB1997,luijtenPRE1999,blote2002}. Both the mean-field and Ising limits are represented by dashed lines in Fig.~\ref{fig02}. The value of the Binder's cumulant at the critical point has a minimum  when the parameter $\alpha$ is assigned to the dimensionality of the underlying square lattice, $\alpha=2$. Moreover, in the range $0 < \alpha < 3$, the $\alpha$-dependent values for the critical cumulant are all located below the mean-field line. This is a rather unusual behavior, since the lower bound for $U^{*}$ normally corresponds to the mean-field value. Curiously, in the framework of the Kleinberg's prescription, at $\alpha=d$, the navigation time has a minimum \cite{kleinberg2000}, 
while the Laplacian transport transport displays a maximum conductance \cite{lucasPRL2014}. The $2D$ Ising behavior is observed only for $\alpha \ge 4$. The results shown in Fig.~\ref{fig02} suggest a crossover between the mean-field and the $2D$ Ising universality classes as one varies the parameter $\alpha$. To investigate the critical behavior of the model, we analyse the finite-size scaling behavior of the system, which allows us to extrapolate the information available from finite-system simulations to the thermodynamic limit. Near the critical point, the finite-size scaling equations for the observables considered here are

\begin{equation}
M_{N}(q) \sim N^{-\beta/\overline{\nu}}\widetilde{M}(\varepsilon N^{1/ \overline{\nu}}),
 \label{eq5}
\end{equation}
\begin{equation}
 \chi_{N}(q) \sim N^{\gamma/\overline{\nu}}\widetilde{\chi}(\varepsilon N^{1/\overline{\nu}}), 
 \label{eq6}
\end{equation}
 \begin{equation}
 U_{N}(q) \sim \widetilde{U}(\varepsilon N^{1/\overline{\nu}}),
 \label{eq7}
\end{equation}
where $\varepsilon=(q-q_{c})$ is the distance from the critical noise. The exponents $\beta$, $\gamma$, and $\overline{\nu}$ are, respectively, associated  to the decay of the order parameter $M_{N}(q)$, the divergence of the susceptibility $\chi_{N}(q)$, and the divergence of the correlation volume ($\xi \sim \varepsilon^{-\overline{\nu}})$. Their exact values for the Ising universality class are $\beta = 1/8$, $\gamma = 7/4$, and $\overline{\nu} = 2$, whereas the mean-field exponents are  $\beta = 1/2$, $\gamma = 1$, and $\overline{\nu} = 2$ \cite{lubeckBook2008}. Notice that we are using $N$ into the definition of the scaling variable $x=\varepsilon N^{1/\overline{\nu}}$, where $\overline{\nu}=d\nu$ \cite{sampaio2013}.

\begin{figure}[t]
\includegraphics*[width=\columnwidth]{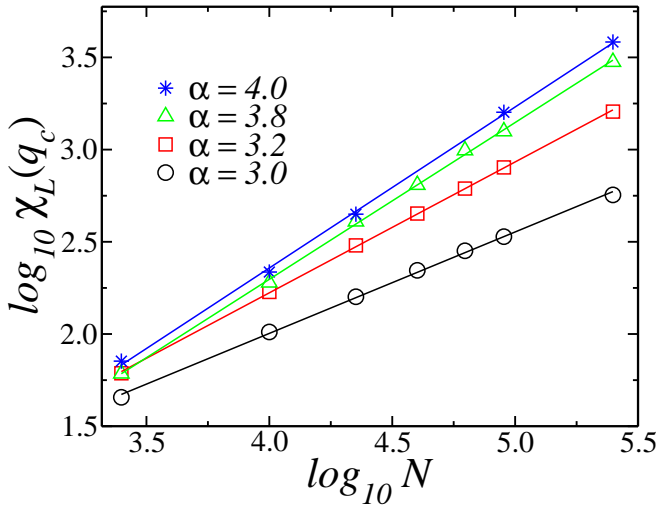}
\caption{(Color online) Logarithmic plot showing the finite-size scaling for the critical susceptibility with $\alpha=3.0$(circles), $3.2$(squares), $3.8$ (triangles), and $4.0$(stars). The solid lines represent the least-squares fits to data, whose slopes corresponds to the exponent $\gamma/\overline{\nu}$ (see Table \ref{table1}).}
\label{fig03}
\end{figure}

\begin{figure}[t]
\includegraphics*[width=\columnwidth]{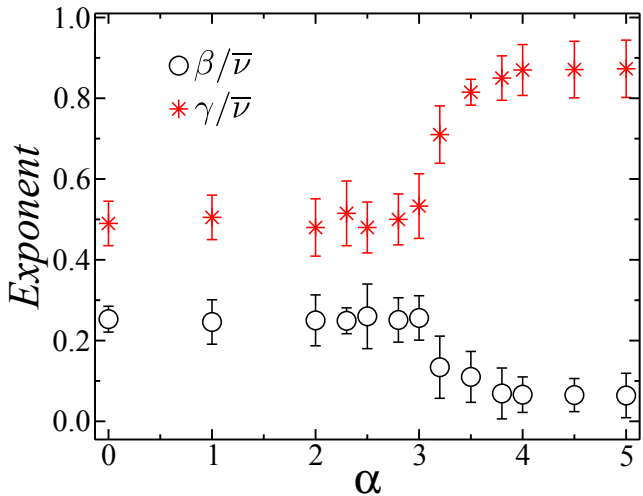}
\caption{(Color online) The dependence of the critical exponents $\beta/\overline{\nu}$ and $\gamma/\overline{\nu}$ with the parameter $\alpha$ (see Table \ref{table1}). Here, $\overline{\nu} = 2$ for all values of $\alpha$.}
\label{fig04}
\end{figure}

\begin{table}
\vspace{3mm}
\caption{\label{table1} Results for the critical noise $q_{c}$ and critical exponents $\beta/\overline{\nu}$ and $\gamma/\overline{\nu}$ of the majority-vote model on Kleinberg's network for different values of the parameter $\alpha$. The mean-field values are $\beta/\overline{\nu}=0.250$ and $\gamma/\overline{\nu} = 0.500$, while the exponents for the two-dimensional Ising model are $\beta/\overline{\nu} = 0.0625$ and $\gamma/\overline{\nu} = 0.875$ \cite{lubeckBook2008,oliveira1992}.}
\begin{ruledtabular}
\begin{tabular}{cccc}
$\alpha$ &$q_{c}$  &$\beta/\overline{\nu}$ &$\gamma/\overline{\nu}$ \\
\hline
 $0.0$      &$0.1998\pm 0.0001$             &$0.253\pm 0.032$            &$0.490\pm 0.055$      \\
 $1.0$      &$0.1997\pm 0.0002$             &$0.25\pm  0.040$            &$0.505\pm 0.055$       \\
 $2.0$      &$0.1892\pm 0.0001$             &$0.250\pm 0.063$            &$0.480\pm 0.060$        \\
 $2.3$      &$0.1754\pm 0.0004$             &$0.249\pm 0.032$            &$0.535\pm 0.071$         \\
 $2.5$      &$0.1650\pm 0.0009$             &$0.260\pm 0.080$           &$0.480\pm  0.063$         \\
 $2.8$      &$0.1464\pm 0.0007$             &$0.251\pm 0.055$            &$0.500\pm 0.063$     \\
 $3.0$      &$0.1350\pm 0.0003$             &$0.256\pm 0.055$            &$0.533\pm 0.050$          \\
 $3.2$      &$0.1235\pm 0.0003$             &$0.134\pm 0.060$            &$0.710\pm 0.070$          \\
 $3.5$      &$0.1097\pm 0.0010$             &$0.110\pm 0.060$            &$0.815\pm 0.032$           \\
 $3.8$      &$0.1005\pm 0.0001$             &$0.069\pm 0.044$            &$0.850\pm 0.055$            \\
 $4.0$      &$0.0963\pm 0.0005$             &$0.066\pm 0.041$            &$0.870\pm 0.060$            \\
 $5.0$      &$0.0820\pm 0.0003$             &$0.064\pm 0.040$            &$0.873\pm 0.065$            \\
\end{tabular}
\end{ruledtabular}
\end{table}

To determine how the parameter $\alpha$ affects the critical behavior of the model, we have explore the dependence of the magnetization and the susceptibility on the system size $N$ at $q=q_{c}$, by considering the finite-size scaling relations Eqs.(~\ref{eq5}) and (~\ref{eq6}). From this analysis, we are able to estimate the exponents $\beta/\overline{\nu}$ and $\gamma/\overline{\nu}$. Moreover, the correlation length exponent $\overline{\nu}$ can be obtained applying the same analysis, but now from the derivative of the Binder's cumulant with respect to the noise parameter.  
For the susceptibility, this analysis is illustrated in Fig.~\ref{fig03} considering four values of the parameter $\alpha$. The results for the critical points and the critical exponents, obtained from simulations with several values of $\alpha$, are summarized in Table~\ref{table1} and Fig.~\ref{fig04}. Within the error bars, we can conclude that, for $0\le \alpha \le 3$, the critical exponents are consistent with those of the mean-field critical behavior, whereas for $\alpha>4$ we get 2D Ising exponents. In the range $3 < \alpha < 4$, we obtain $\alpha$-dependent exponents, so that the critical behavior of the majority-vote model is neither described by mean-field nor by Ising universality classes. This continuous variation of the critical exponents with $\alpha$ is consistent with the observed effective dimensionality in spatially embedded networks \cite{havlinNatPhys2011}. 

\begin{figure}[t]
\includegraphics*[width=\columnwidth]{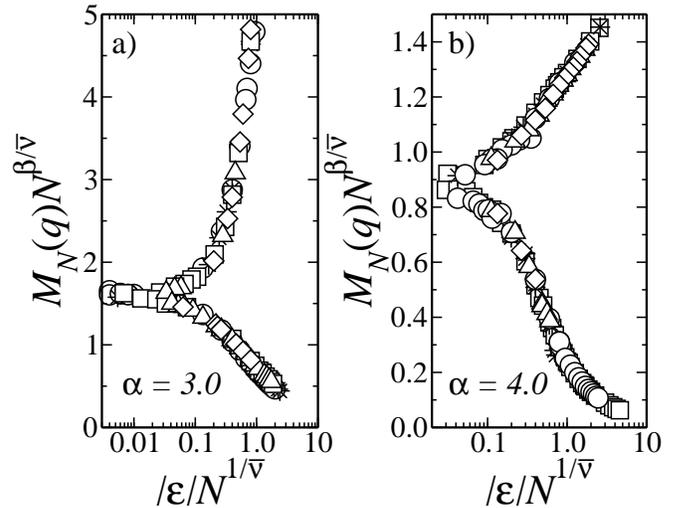}
\caption{Data collapse for the magnetization for system sizes $N=10000 \mbox{ (circles)}$, $22500 \mbox{ (stars)}$, $40000 \mbox{ (rectangles)}$, $62500 \mbox{ (triangles)}$, and $90000 \mbox{ (diamonds)}$. The universal curve for $\alpha=3$ is consistent with mean-field exponents: $\beta = 1/2$,  $\gamma = 1$,  $\overline{\nu} = 2$. For $\alpha=4$, the data collapse is obtained using Ising exponents: $\beta = 1/8$,  $\gamma = 7/4$, $\overline{\nu} = 2$.}
\label{fig05}
\end{figure}

In order to accurately determine the exponents and the nature of the continuous phase transition, we now consider the data collapse of the results from our simulations with different system sizes $N$ for a fixed value of $\alpha$. These data collapses reflect (see Eqs.(\ref{eq5}) and (\ref{eq6})) the existence of universal functions for the re-scaled magnetization $\widetilde{M}(x) = M_{N}(q)N^{\beta/\overline{\nu}} $ and for the re-scaled susceptibility $\widetilde{\chi}(x) = \chi_{N}(q)N^{-\gamma/\overline{\nu}}$, with both depending only on the scaling variable $x = \varepsilon N^{1/\overline{\nu}}$. The universal curves shown in Fig.~\ref{fig05} for the order parameter (magnetization) reveal the presence of two regimes. In Fig.~\ref{fig05}(a) the resulting data collapse is compatible with mean-field critical behavior, that is, the universal function for $\alpha=3$ is consistent with mean-field exponents: $\beta = 1/2$,  $\gamma = 1$, and $\overline{\nu} = 2$. The same set of exponents were considered to obtain data collapses of excellent quality for others values of $\alpha < 3$. However, the data collapse for $\alpha=4$ shown in Fig.~\ref{fig05}(b) is consistent with $2D$ Ising exponents. Satisfactory data collapse were also obtained for the susceptibility and the Binder's cumulant (not shown). 

\begin{figure}[t]
\includegraphics*[width=\columnwidth]{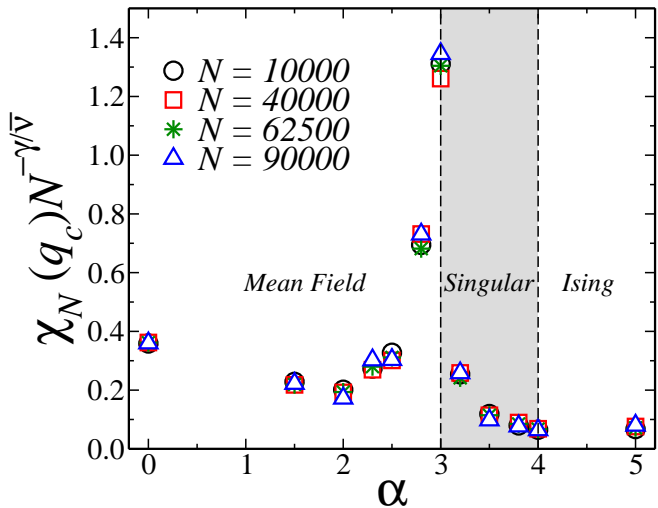}
\caption{(Color online) Data collapse for the susceptibility at the critical point as a function of $\alpha$. The exponents used are those from our conjecture: for $\alpha \le 3$, $\gamma/\overline{\nu} = 0.5$ (mean-field), for $3<\alpha < 4$ the exponents are $\alpha$-dependent (see Table~\ref{table1}), and for $\alpha\ge 4$,  $\gamma/\overline{\nu} = 0.875$ (2D Ising). Notice that the maximum of the fluctuations occurs at $\alpha = 3$.}
\label{fig06}
\end{figure}

Figure~\ref{fig06} shows the data collapse for the critical amplitude \cite{lubeckPRL2003,sampaio2013} of the susceptibility as a function of the parameter $\alpha$, considering four values of system sizes $N= 10000, 40000, 62500$, and $90000$. More precisely, we plot the re-scaled susceptibility, $\chi_{N}(q_{c})N^{-\gamma/\overline{\nu}}$, using the set of calculated exponents, namely, mean-field exponents for $\alpha<3$, $\alpha$-dependent exponents for $3 < \alpha < 4$, and 2D Ising exponents for $\alpha>4$. As can be seen, the results shown in Fig.~\ref{fig06} give support to our conjecture, highlighting the singular region, characterized by non-universal exponents. Moreover, $\alpha = 2$ is not associated with the lowest limit for the critical amplitude of the susceptibility, differently from the critical amplitude of the Binder's cumulant. However, an upper limit can be observed for $\alpha=3$ in the fluctuations of the order parameter.

\section{\label{sec:level4}Conclusions}

In this work, the effects of nonlocal interactions on the phase diagram and critical behavior of the majority-vote model on Kleinberg's networks are determined by Monte Carlo simulations and finite-size scaling analysis. The model is defined in terms of the noise parameter $q$ associated with the resistance for accepting the majority state and the control parameter $\alpha$ for the addition of long-range connections (shortcuts). The resulting phase diagram in the $\alpha$ vs $q$ parameter space indicates that the critical noise, $q_c\left(\alpha\right)$, above which the system does not display global order (consensus) decreases with $\alpha$. The Binder's cumulant calculated at the critical noise, whose value has been usually considered as an indicative of a given class of universality, yields results below the mean-field line as $\alpha$ varies in the interval $0< \alpha <3$, where a minimum occurs at $\alpha = 2$. Nevertheless, for the entire region $0 \le \alpha \le 3$ the obtained set of critical exponents is consistent with mean-field behavior. On the other hand, for $\alpha\ge 4$, the calculated values of critical Binder's cumulant and critical exponents are both indicative of a system that belongs to the two-dimensional Ising universality class. Finally, in the region $3 < \alpha < 4$, a continuum crossover can be observed from mean-field to Ising critical behavior, which  suggests that the majority-vote model on Kleinberg’s networks is described by $\alpha$-dependent exponents.

\begin{acknowledgments}
We thank the Brazilian agencies CNPq, CAPES, FUNCAP, and the National Institute of Science and Technology for Complex Systems for financial support.
\end{acknowledgments}

\end{document}